# SUPPORTING INFORMATION FOR:
# Enthalpy of formation for Cu-Zn-Sn-S (CZTS)

## Sergey V. Baryshev and Elijah Thimsen

### Experimental sputtering rates for MgO and ZnO.

From a previously published work,[1] we had data on the relative sputtering rate of well-defined nanolayers of MgO and ZnO at orthogonal ion incidence. We note that the ion current ($I$) and sputtered crater area ($A$) used in that work were different than those used for the CZTS samples discussed in the main text, which gives rise to a different absolute value for the ZnO sputtering rate – essentially, this is because $J=I/A$ was about 3 times smaller. A multilayer consisting of alternating layers of ZnO, 6.4 nm in thickness, and MgO, 6.1 nm in thickness, was deposited by atomic layer deposition. The measured SIMS profiles, acquired under low energy (250 eV) $Ar^+$ bombardment are presented below.

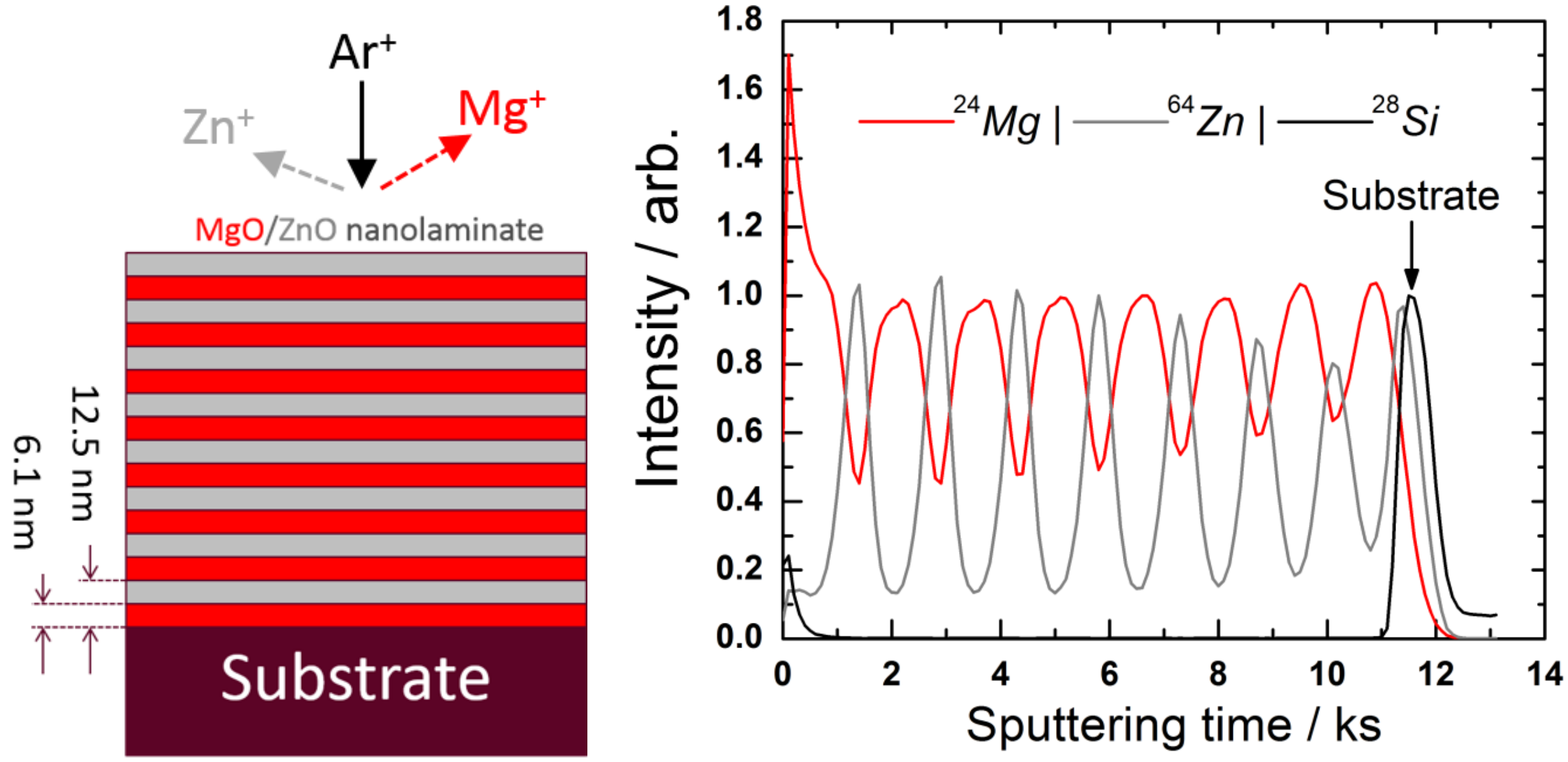

**Figure S1. TOF-SIMS depth profiled of a periodic multilayer consisting of alternating layers of ZnO (6.4 nm) and MgO (6.1 nm).**

Since the thicknesses of the layers in our nanolaminate are known (Figure S1), we need only determine the time it takes to sputter through each layer to get the sputtering rate for each material. From Figure S1, it can be seen that the profile consists of periodic alternating peaks of Mg and Zn. The full width at half maximum (FWHM) is taken as the time it takes to sputter though a given layer. The FWHM for the Zn was measured by averaging the FWHM of the first three peaks, determined by fitting three Gaussian functions (Figure S2). The average FWHM from the first three Zn peaks is 521 seconds, yielding a sputtering rate of $SR_{ZnO}$=0.012 nm/s.

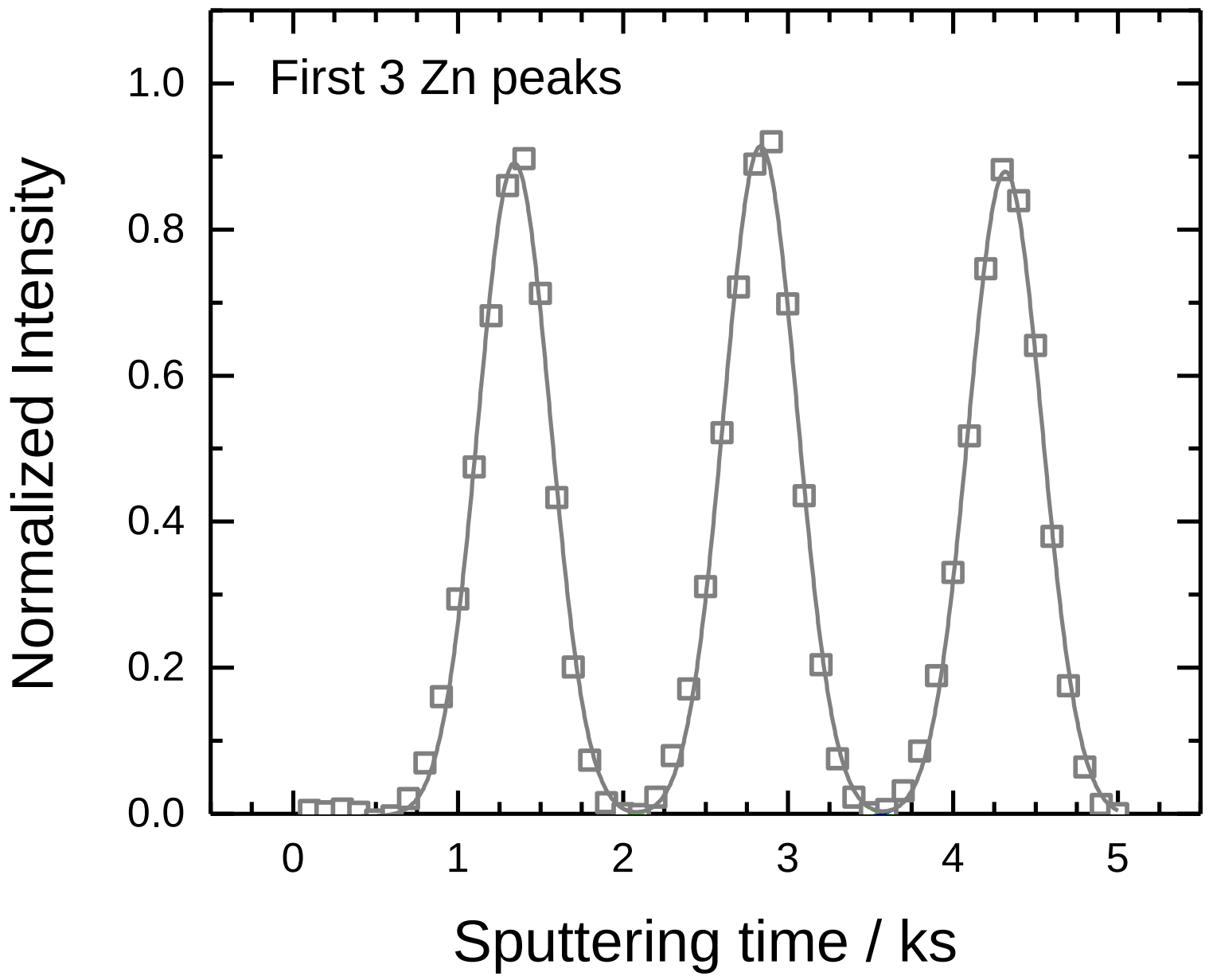

**Figure S2. Gaussian fit used to determine the FWHM of the first three Zn peaks in the SIMS profile of Figure S1. The spectrum was baseline-subtracted to obtain a more accurate fit.**

A slightly different procedure was used for the Mg profile. Since the peaks had relatively flat summits, the profile was baseline subtracted, normalized and the FWHM was measured manually from the plots at the 0.5 intensity line. The average of the FWHM measured from the 2$^{nd}$, 3$^{rd}$ and 4$^{th}$ Mg peaks was 996 seconds (Figure S3), yielding an average sputtering rate of 0.0061 nm/s. We pause to note that the sum of the two FWHM values for ZnO and MgO (1517 seconds) is in excellent agreement with the peak-to-peak spacing (1507 seconds), both of which correspond to sputtering through one MgO layer and one ZnO layer. And thus, the experimental sputtering rate ratio is

(S1) $$\left(\frac{SR_{ZnO}}{SR_{MgO}}\right)_{exp} = 2.0$$

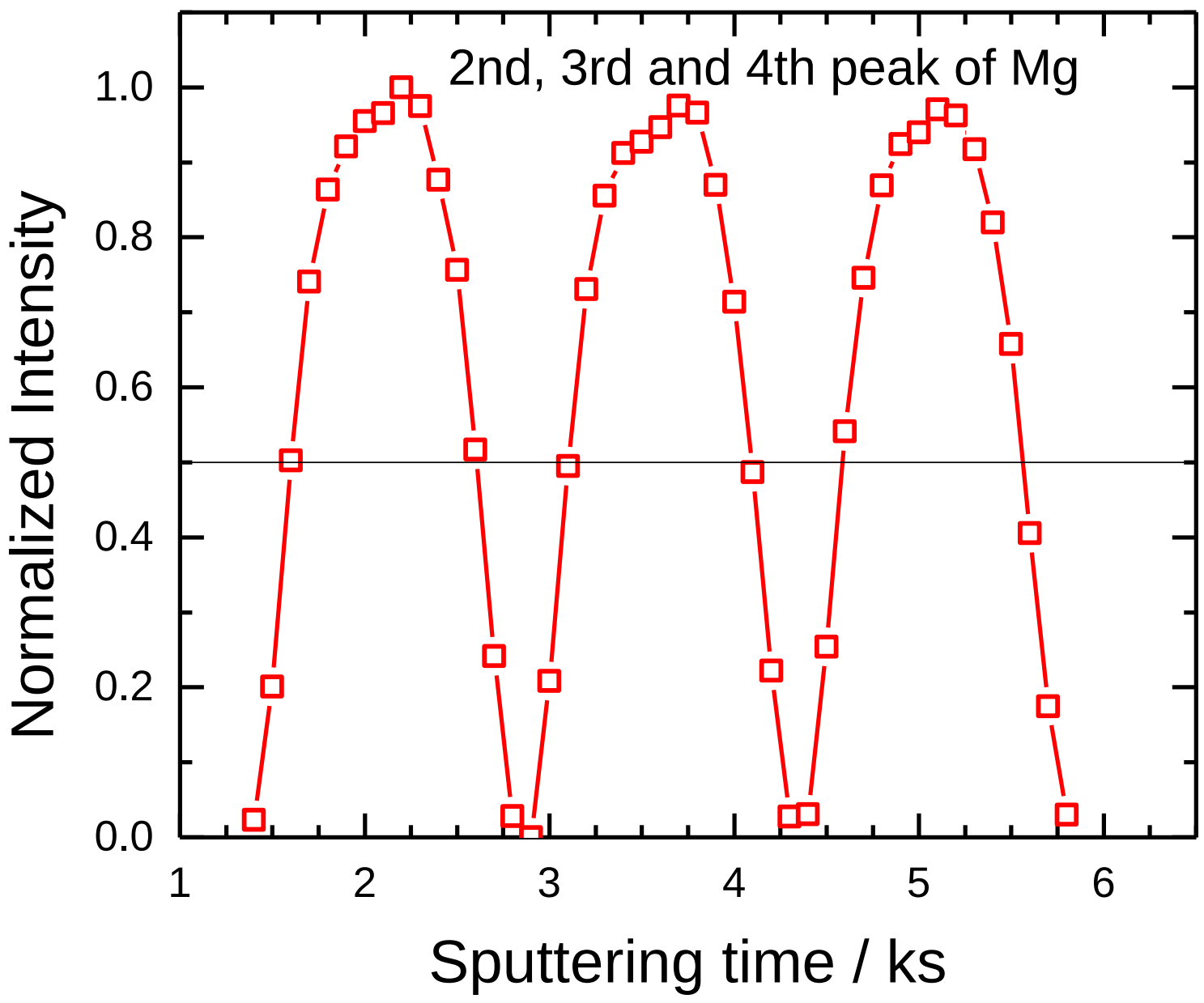

**Figure S3. Baseline-subtracted, normalized Mg peaks used to determine the average FWHM of the SIMS profile in Figure S1.**

**Experimental sputtering rates of ZnO, ZnS and CZTS from the measured sputtering profiles.**

Since the thicknesses of the ZnO, ZnS and CZTS layers are known, we need to extract the time at which the ZnO/CZTS, ZnS/CZTS, and CZTS/Si substrate interfaces are reached to extract the time it takes to sputter through each layer. The times at which the ZnO/CZTS and ZnS/CZTS interfaces were reached were taken as the time at which the Zn intensity was halfway between the intensity in the CZTS layer and the intensity in the ZnO or ZnS layer. The determination of the time at which the ZnO/CZTS and ZnS/CZTS interfaces were reached is summarized in Figure S4. The time at which the CZTS/Si substrate interface was reached was determined by the peak position of the $Si^{28}$ signal, since this signal peaks because of the matrix effect of the native oxide (see Figure 1 of main text). The sputtering rates determined using this procedure are summarized as follows:

$$SR_{ZnS} = 0.065 \; nm\, s^{-1}$$
$$SR_{ZnO} = 0.035 \; nm\, s^{-1}$$
$$SR_{CZTS} = 0.061 \, nm\, s^{-1}$$
$$\left(\frac{SR_{ZnS}}{SR_{ZnO}}\right)_{exp} = 1.9 \qquad \text{(S2)}$$
$$\left(\frac{SR_{ZnS}}{SR_{CZTS}}\right)_{exp} = 1.1$$
$$\left(\frac{SR_{ZnO}}{SR_{CZTS}}\right)_{exp} = 0.57$$

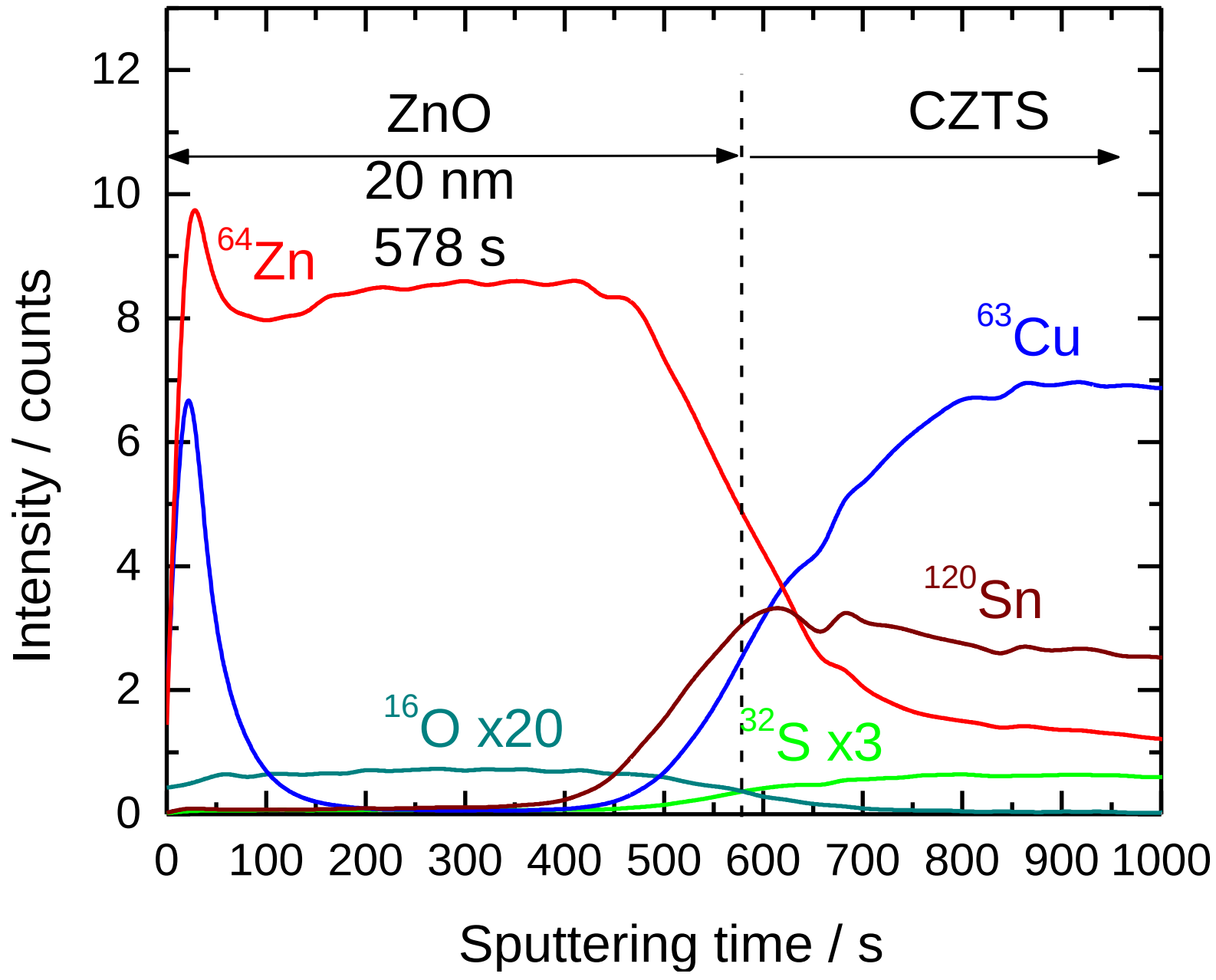

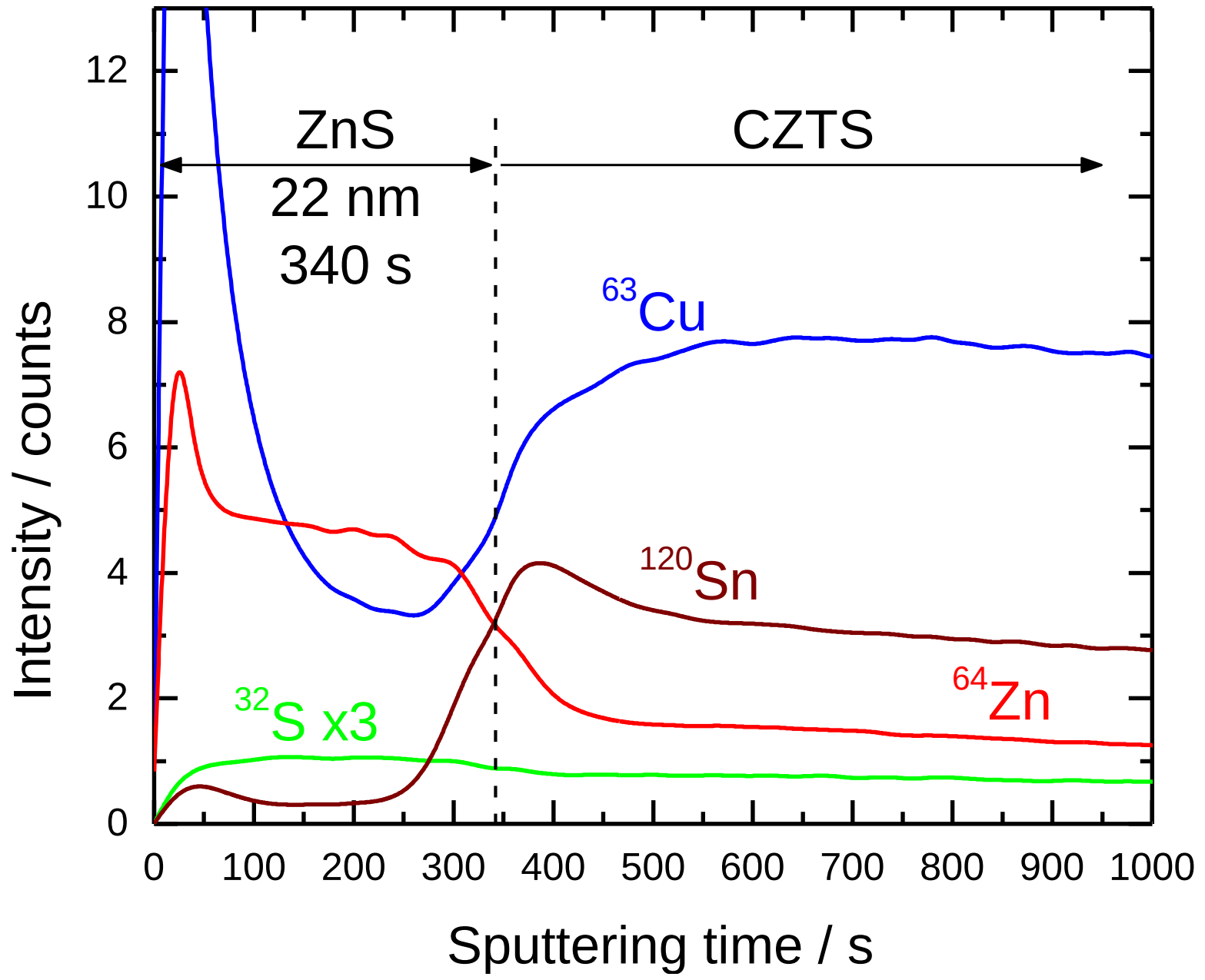

**Figure S4. Expanded plots showing the determination of the ZnO/CZTS interface (top) and ZnS/CZTS interface (bottom).**

**Comparison of the experimental values to the model predictions.**

The surface binding energy $U_0$ for a material can be calculated using the Born-Haber cycle (Figure S5).

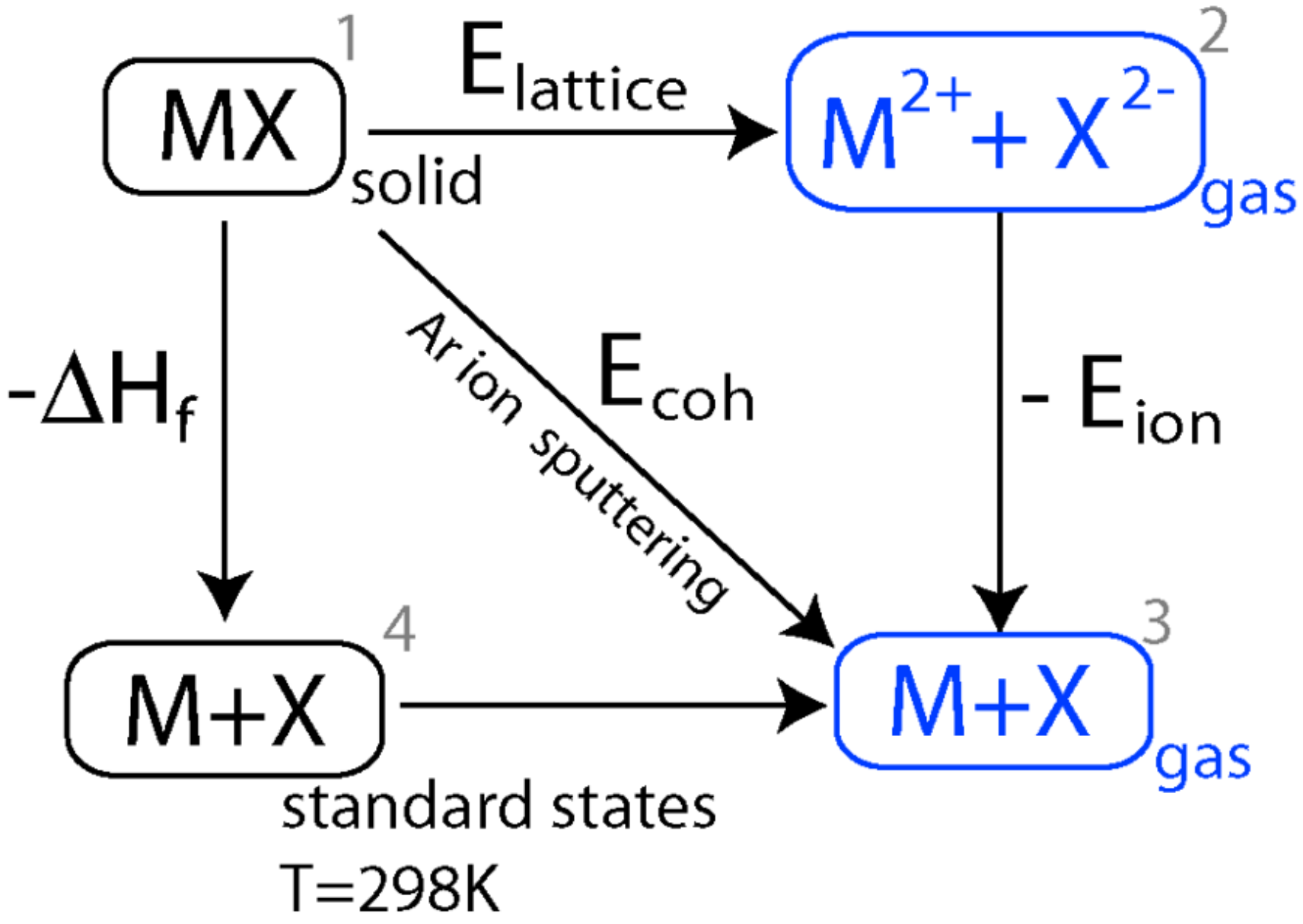

**Figure S5. Schematic of the Born-Haber cycle for MgO, ZnO and ZnS.**

Rearranging equation 9 from the main text, we can write an expression for the surface binding energy $U_0$ as a function of tabulated thermodynamic values for the known materials MgO, ZnO and ZnS:

$$\text{(S3)} \qquad U_0 = \frac{E_{lattice} - E_{ion}}{\sum \nu_i} = \frac{E_{lattice} - \left[ \sum_{cations} \left( \nu_{M_i} \cdot \sum_{j=1}^{n} I(M_i^{j+}) \right) + \sum_{anions} \left( \nu_{A_i} \cdot \sum_{j=1}^{m} I(A_i^{j-}) \right) \right]}{\sum \nu_i}.$$

For MgO and ZnO, equation (S3) simplifies to:

$$\text{(S4)} \qquad U_0 = \frac{E_{lattice} - E_{ion}}{\sum \nu_i} = \frac{E_{lattice} - \left[ I(M_i^{+}) + I(M_i^{2+}) + I(O^{-}) + I(O^{2-}) \right]}{2}.$$

Where $M_i$ is either Mg or Zn. For ZnS, equation S3 simplifies to:

$$\text{(S5)} \qquad U_0 = \frac{E_{lattice} - E_{ion}}{\sum \nu_i} = \frac{E_{lattice} - \left[ I(Zn^{+}) + I(Zn^{2+}) + I(S^{-}) + I(S^{2-}) \right]}{2}.$$

Using the values in Table S1, the following relative sputtering rates are predicted using equation 8 from the main text:

$$\text{(S6)} \qquad \begin{aligned} \left( \frac{SR_{ZnO}}{SR_{MgO}} \right)_{theory} &= \frac{n_{MgO}}{n_{ZnO}} \cdot \frac{\alpha_{ZnO}}{\alpha_{MgO}} \cdot \frac{\gamma_{ZnO}}{\gamma_{MgO}} \cdot \frac{U_{0,MgO}}{U_{0,ZnO}} = 2.1 \\ \left( \frac{SR_{ZnS}}{SR_{ZnO}} \right)_{theory} &= \frac{n_{ZnO}}{n_{ZnS}} \cdot \frac{\alpha_{ZnS}}{\alpha_{ZnO}} \cdot \frac{\gamma_{ZnS}}{\gamma_{ZnO}} \cdot \frac{U_{0,ZnO}}{U_{0,ZnS}} = 2.1 \end{aligned}$$

These predicted values are in excellent agreement with the experimental measurements present above in equations S1 and S2.

| **Table S1: Summary of parameters. Values taken from the CRC Handbook of Chemistry and Physics unless otherwise stated.** | | |
|---|---|---|
| Parameter | Value | Units |
| $2U_0(ZnO)=E_{coh}(ZnO)$ | 8.3 or 7.5[a)] (average 7.9) | eV/molecule |
| $2U_0(MgO)=E_{coh}(MgO)$ | 9.4 | eV/molecule |
| $2U_0(ZnS)=E_{coh}(ZnS)$ | 6.3[a)] | eV/molecule |
| $E_{lattice}(ZnO)$ | 42.93 or 42.06[b)] (average 42.5) | eV/molecule |
| $E_{lattice}(MgO)$ | 39.33 | eV/molecule |
| $E_{lattice}(ZnS)$ | 36.95[b)] | eV/molecule |
| $I(O^-)$ | -1.47 | eV |
| $I(O^{2-})$ | 8.75 | eV |
| $I(S^-)$ | -2.07 | eV |
| $I(S^{2-})$ | 5.51 | eV |
| $I(Zn^+)$ | 9.39 | eV |
| $I(Zn^{2+})$ | 17.96 | eV |
| $I(Mg^+)$ | 7.64 | eV |
| $I(Mg^{2+})$ | 15.04 | eV |
| $\alpha$(ZnO, Ar) | 0.26 | – |
| $\alpha$(MgO, Ar) | 0.21 | – |
| $\alpha$(ZnS, Ar) | 0.27 | – |
| $\alpha$(CZTS, Ar) | 0.29 | – |
| $\gamma$(ZnO, Ar) | 1.0 | – |
| $\gamma$(MgO, Ar) | 0.89 | – |
| $\gamma$(ZnS, Ar) | 0.99 | – |
| $\gamma$(CZTS, Ar) | 0.98 | – |
| n(ZnO) | $8.3\times10^{22}$ | $cm^{-3}$ |
| n(MgO) | $10.7\times10^{22}$ | $cm^{-3}$ |
| n(ZnS) | $5.1\times10^{22}$ | $cm^{-3}$ |
| n(CZTS) | $5.1\times10^{22}$ | $cm^{-3}$ |

a) Calculated using lattice energies from classical work [2].

b) Taken from classical work [2].

**Copper zinc tin sulfide.**

Having now vetted the procedure for relating sputtering rates in the low energy regime to the surface binding energy for samples of known composition, we can move on to measuring the surface binding energy for copper zinc tin sulfide. The ideal stoichiometry of the photovoltaic absorber known as CZTS is $Cu_2ZnSnS_4$. In practice the best performance of this material in photovoltaic solar cells is achieved in samples that are copper poor and zinc rich.[3] An interesting feature of the quaternary system is that it can accommodate a wide variety of metal ratios while still crystallizing in the same crystal structure.[4] It is believed that this compositional flexibility is due to close-packed sulfur sublattice. We measured the sputtering rate of 3 samples of CZTS that had the same thickness and composition. The composition was measured by x-ray fluorescence to be $Cu_{1.9}Zn_{1.5}Sn_{0.8}S_4$. The film thickness was 92 nm. The average sputtering rate of these three samples was 0.061 nm/s with a standard deviation of 6%.

Recall that in the present model, multicomponent materials with many different types of atoms are treated as materials with the same density of atoms, but all of those model atoms are the same and have the number-average atomic mass of the real multicomponent material. Therefore one must know the composition. For our CZTS sample, we calculate a target mass of 53.91 g mol$^{-1}$ by the following equation

(S7) $$M_t = \frac{\nu_{Cu} \cdot M_{Cu} + \nu_{Zn} \cdot M_{Zn} + \nu_{Sn} \cdot M_{Sn} + \nu_S \cdot M_S}{\nu_{Cu} + \nu_{Zn} + \nu_{Sn} + \nu_S},$$

where $\nu_i$ is the number of atoms of species i in the molecule (e.g. 1.9 for Cu) and $M_i$ is the mass of species i. To calculate the atomic density, n, a mass density for CZTS of 4.6 g/cm$^3$ was used. Then the atomic density can be calculated to be $5.1\times10^{22}$ atoms per cm$^3$. The parameters α and γ can be calculated using equations 6 and 7 of the main text.

Equation 8 of the main text can now be rearranged to calculate the surface binding energy of CZTS from the relative sputtering rates and compositional parameters

(S8) $$U_{0,CZTS} = U_{0,ZnO} \cdot \frac{n_{ZnO}}{n_{CZTS}} \cdot \frac{\alpha_{CZTS}}{\alpha_{ZnO}} \cdot \frac{\gamma_{CZTS}}{\gamma_{ZnO}} \cdot \frac{SR_{ZnO}}{SR_{CZTS}} = 4.0\, eV$$
$$U_{0,CZTS} = U_{0,ZnS} \cdot \frac{n_{ZnS}}{n_{CZTS}} \cdot \frac{\alpha_{CZTS}}{\alpha_{ZnS}} \cdot \frac{\gamma_{CZTS}}{\gamma_{ZnS}} \cdot \frac{SR_{ZnS}}{SR_{CZTS}} = 3.6\, eV$$

by averaging these values, we get a value for $U_{0,CZTS}$ = 3.8±0.4 eV. Note that our model assumes a single type of atom with an atomic weight equal to the number average of all of the elements in CZTS. Therefore the surface binding energy is for one atom, not one molecule of CZTS.

We may now solve equation 9 of the main text for the standard enthalpy of formation as a function of the measured surface binding energy, number of atoms in the molecule and tabulated thermodynamic data (Table S2):

$$-\Delta H_f^0 = U_0 \sum \nu_i - \left( \sum \nu_i \cdot \Delta H_{vap,i} + \sum_{anions} \nu_i \eta_i D^0_{molecule,i} \right)$$

(S9)

$$= 3.8\frac{eV}{atom} \times 8.2\frac{atom}{molecule} - \begin{pmatrix} 1.9\frac{atom}{molecule} \times 3.11\frac{eV}{atom} + \\ 1.5\frac{atom}{molecule} \times 1.20\frac{eV}{atom} + \\ 0.8\frac{atom}{molecule} \times 3.01\frac{eV}{atom} + \\ 4\frac{atom}{molecule} \times 0.12\frac{eV}{atom} + \\ 4\frac{atom}{molecule} \times 1\frac{bond}{atom} \times 2.75\frac{eV}{bond} \end{pmatrix} = 9.6\frac{eV}{molecule}$$

The standard enthalpy of formation calculated using equation S9 can also be rewritten as -1.2 eV atom$^{-1}$; or -930 kJ mol$^{-1}$.

**Table S2. Summary of parameters used in the calculation of the enthalpy of formation for CZTS using equation S9.**

| Parameter | Value | Units |
|---|---|---|
| $\Delta H_{vap}$ (Zn) | 1.20[a)] | eV atom$^{-1}$ |
| $\Delta H_{vap}$ (Sn) | 3.01[a)] | eV atom$^{-1}$ |
| $\Delta H_{vap}$ (Cu) | 3.11[a)] | eV atom$^{-1}$ |
| $\Delta H_{vap}$ (S) | 0.12[b)] | eV atom$^{-1}$ |
| $D^0$(S-S) | 2.75[c)] | eV bond$^{-1}$ |

a) Taken from [5].
b) Taken from [5] and [6].
c) Taken from [7].

**Calculation of the free energy of formation for the binary metal sulfides.**

| **Table S3. Reactions and data used to calculate $\Delta G_R$ in the main text. The expressions for the energy of formation for the binary sulfides were taken from [8].** | | | |
|---|---|---|---|
| Index | Reaction | ΔG/kJ mol$^{-1}$ (T in K) | T range/$^o$C |
| **R1** | $Cu_2S + ZnS + SnS_2 \leftrightarrow Cu_2ZnSnS_4$ | $\Delta G_r = \Delta G_{f,CZTS} - \Delta G_{f,binaries}$ | |
| **R2** | $4Cu + S_2 \leftrightarrow 2Cu_2S$ | $\Delta G_{R2} = -268 + 0.072\ T$ | 103 – 425 |
| | | $\Delta G_{R2} = -260 + 0.061\ T$ | 435 – 1067 |
| **R3** | $2Zn + S_2 \leftrightarrow 2ZnS$ | $\Delta G_{R3} = -538 + 0.191\ T$ | 25 – 420 |
| | | $\Delta G_{R3} = -549 + 0.207\ T$ | 420 – 1200 |
| **R4** | $2Sn + S_2 \leftrightarrow 2SnS$ | $\Delta G_{R4} = -344 + 0.173\ T$ | 25 – 232 |
| | | $\Delta G_{R4} = -354 + 0.193\ T$ | 232 – 600 |
| **R5** | $4SnS + S_2 \leftrightarrow 2Sn_2S_3$ | $\Delta G_{R5} = -234 + 0.200\ T$ | 25 – 600 |
| **R6** | $2Sn_2S_3 + S_2 \leftrightarrow 4SnS_2$ | $\Delta G_{R6} = -234 + 0.200\ T$ | 25 – 760 |

From reactions R2-R6, the free energy of formation for the binary metal sulfides is straightforward to calculate by considering the reaction stoichiometry:

(S10) $$\Delta G_{f,binaries} = \frac{\Delta G_{R2} + \Delta G_{R3}}{2} + \frac{2\Delta G_{R4} + \Delta G_{R5} + \Delta G_{R4}}{4}$$

| **Table S4: Summary of enthalpies of formation reported for CZTS by various authors as well as the constituent binary compounds** | | |
|---|---|---|
| Material | $\Delta H_f^0$/kJ mol$^{-1}$ | Reference |
| CZTS | -903 | This work |
| CZTS | -337 | Maeda et al.[9] |
| CZTS | -406 | Walsh et al.[10] |
| $Cu_2S$+ZnS+$SnS_2$ | -438 | |
| $Cu_2S$ | -79.5 | Kubachewski et al.[11] |
| ZnS | -205 | Kubachewski et al.[11] |
| $SnS_2$ | -153 | Kubachewski et al.[11] |